\documentclass[final,1p,times]{elsarticle}
\usepackage{amssymb}
\usepackage{graphicx}

\newcommand{\mib}{$\mu_{B}$ }
\newcommand{\ptopi}{$p/\rm \pi$ }

\newcommand{\ptopip}{$p/\rm \pi^{+}$ }
\newcommand{\ptopipn}{$p/\rm \pi^{+}$}
\newcommand{\aptopi}{$\bar{p}/\rm \pi^{-}$ }

\newcommand{\rootsnn}[1]{$\sqrt{s_{NN}} = #1$~GeV}
\newcommand{\roots}[1]{$\sqrt{s} = #1$~GeV}

\newcommand{\pt}{$p_{T}$ }
\newcommand{\ptn}{$p_{T}$}
\newcommand{\pteq}[1]{$p_{T} = #1$~GeV/c}

\newcommand{\ptapp}[1]{$p_{T} \approx #1$~GeV/c}

\journal{Nuclear Physics A} 
\begin{document}
\begin{frontmatter}

\title{The rapidity dependence of the proton-to-pion ratio in Au+Au and p+p collisions at \rootsnn{62.4} and 200 GeV}
\author{P. Staszel$^{a}$ for the BRAHMS collaboration}

\address[a]{Smoluchowski Inst. of Physics, 
  Jagiellonian University, 
  ul. Reymonta 4, 
  30-059 Krak\'ow, Poland}

\begin{abstract}

The BRAHMS measured proton-to-pion ratios in Au+Au and
p+p collisions at \rootsnn{62.4} and \rootsnn{200}
are presented as a function of transverse momentum and collision centrality
within the pseudo-rapidity range 0 $\leq\eta \leq$ 3.8 
The results for Au+Au at \rootsnn{200} are compared with predictions
from models which incorporate hydro-dynamics, hadron rescattering and
jet production, in the $\eta$ interval covered.
In Au+Au collisions at \rootsnn{200}, $\eta \approx 2.2$,
and at \rootsnn{62.4}, $\eta = 0$, the bulk medium can be
characterized by the common value of \mib $\approx 65$ MeV.
The  \ptopipn $(p_T)$ ratios measured for these two selections
display a striking agreement in the \pt range covered (up to $2.2$ GeV/c).
At a collision energy of 62.4 GeV and at forward pseudo-rapidity we found a
crossing point of \ptopip ratios measured in central and semi-peripheral
Au+Au and in p+p reactions. The crossing occurs in the narrow $\eta$
bin around value of 3.2, simultaneously in the whole covered \pt range
(0.3 GeV/c $< p_{T} <$ 1.8 GeV/c).

\end{abstract}

\end{frontmatter}

The measured \pt dependence of the baryon-to-meson ratio appears to be 
related   to modifications in  the hadronization mechanisms as it happens in a partonic medium.
It was pointed out that the baryon-to-meson ratio \pt dependence should be 
sensitive to the hadronization scenario due to the different quark content of baryons and
mesons \cite{rj_fries} and/or to radial flow of the bulk medium because of significant
differences in baryon and meson masses \cite{peitzmann}. 
Both flow and medium quark coalescence are expected to 
enhance protons over pions at intermediate \ptn.



The PHENIX \aptopi data at mid-rapidity 
is well described by the  Greco, Ko, and Levai quark coalescence model where the introduced
coalescence involves partons from the medium (thermal) and partons from mini-jets \cite{GKL}.
The Hwa and Yang quark recombination model  is also successful 
in describing BRAHMS and PHENIX mid-rapidity data for \ptopip \cite{hwa_67}.
  
On the other hand, the comparison with the hydrodynamical model shows that hydro-flow
cannot itself account for the large observed ratio above $\approx 3$ GeV/c and that 
the model overpredicts the data at low \pt \cite{ejkm}. 
These results support the view of a hadronization process driven by 
parton recombination with negligible final state interactions
between produced hadrons.
Nevertheless, at large $\mu_{B}$, a significant gap
between the temperature of the transition from the partonic to the
hadronic phase, $T_{c}$, and the temperature of chemical freeze-out is
predicted by QCD lattice calculations \cite{LQCD_vs_freeze-out}. Thus at large $\mu_{B}$, the
picture, suggested by mid-rapidity measurements, 
might be contaminated by final state hadron interactions leading to 
a transition from the parton recombination scheme to
a hydrodynamical description that has a common velocity field for
baryons and mesons \cite{hirano, broniowski_flor}. 

 
The setup of the BRAHMS experiment is described in details
in \cite{brahms_det}. Here we just point out that 
the arrangement of BRAHMS spectrometers, namely of the Mid-Rapidity
Spectrometer (MRS) and the Forward Spectrometer (FS), makes it possible
to measure identified particle spectra over a pseudo-rapidity interval
from $\eta = 0$ to $\eta = 3.8$. 
Particle identification in the FS 
is provided by TOF measurements for low and medium particle momenta.
High momentum particles are identified using
a Ring Imaging Cherenkov detector (RICH) \cite{rich_nim}.

The data analysis utilizes the feature of the same pion and proton
acceptance in the $\eta$ versus \pt space in the same real time measurement.
For a given $\eta$-\pt bin the \ptopi ratios are calculated on a
setting by setting basis. In order to avoid mixing different PID techniques,
which usually lead to different systematic uncertainties,  
the ratios are calculated separately for the TOF PID and the RICH PID. 
In this way all factors such as acceptance corrections, 
tracking efficiencies, trigger normalization and bias related to the centrality cut
cancel out in the ratio. 
The only remaining corrections are those that are species related which are: 
\begin{enumerate}[(i)]
\item decay in flight, interaction with the beam pipe,
and the detector material budget,\label{enu1}
\item the PID efficiency correction.
\end{enumerate} 
The corrections for (\ref{enu1}) are determined from 
the single particle response of pions and protons in
a realistic GEANT \cite{GEANT} model description of the BRAHMS experimental setup.
We estimate that the overall systematic uncertainty related to
this correction  is at the level of $2\%$.  
The TOF PID is done separately for small momentum bins by fitting a
multi-Gaussian function to the experimental squared mass $M^{2}$
distribution and applying a $\pm 3 \sigma$ cut to select a given
particle type. For measurements done with the FS spectrometer 
in the momentum range where pions overlap with kaons,
(usually above $3.5$ GeV/c) the RICH detector can be used in veto mode  
to select kaons  with momentum smaller than  the kaon Cherenkov threshold
which is about $9$ GeV/c. Above the proton threshold momentum, which is
about $15$ GeV/c, the RICH provides a direct proton
identification. In this momentum range the RICH PID is based on the particle separation in the
$M^{2}$ versus momentum space. 
The described PID procedure leads to a relatively clean sample of pions with some
contamination by kaons having spurious rings associated in the RICH counter.
Together with the kaon - proton overlap at larger momenta, this contamination
effect is a source of systematic errors which have been estimated. 



\begin{figure}[t]
\begin{minipage}[t]{68mm}
\centering
\includegraphics[width=1.02\textwidth,totalheight=0.34\textheight]
                {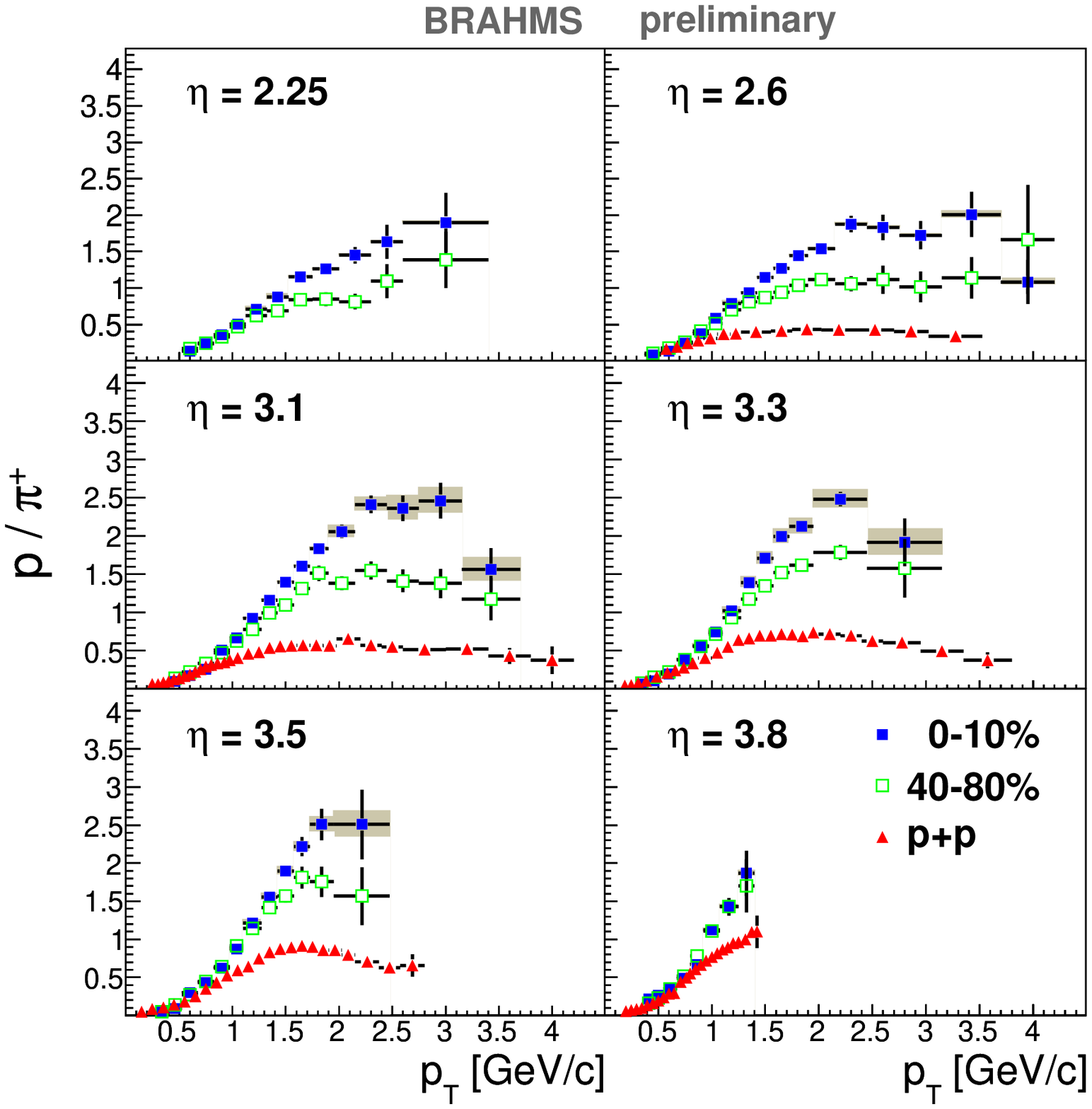}
\end{minipage}
\hspace{\fill}
\begin{minipage}[t]{68mm}
\centering
\includegraphics[width=1.02\textwidth,totalheight=0.34\textheight]
                {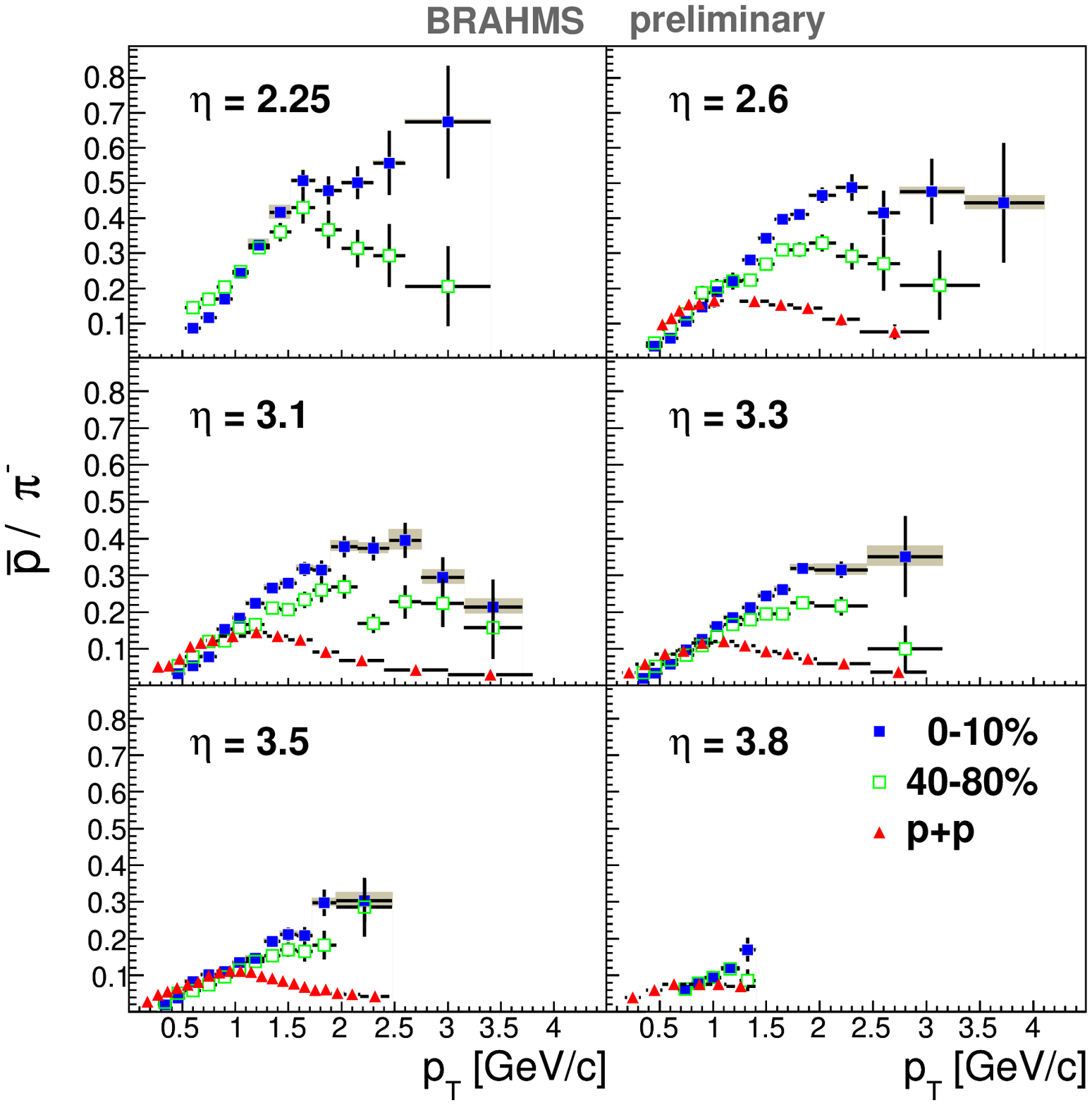}

\end{minipage}

\vspace{-0.5cm}

\begin{minipage}[t]{135mm}
  \caption{Centrality dependent \ptopip (left) and \aptopi (right) ratio for Au+Au system  
    colliding at \rootsnn{200} for central ($0-10\%$) and  
    semi-peripheral ($40-80\%$) reactions in comparison with 
    p+p collision data at the same energy. The vertical bars represent the  
    statistical errors and the shaded bands (plotted only for central
    Au+Au) show the systematic uncertainties.} 
  \label{fig1_2}
\end{minipage}
\end{figure}

Figure \ref{fig1_2} (left) shows \ptopip ratios obtained for Au+Au reacting at \rootsnn{200} 
for two centrality sets of events,
namely, for centrality $0-10\%$ (solid dots) and $40-80\%$ (open
squares) 
The shaded boxes plotted for the most central events only,
represent the systematic uncertainties discussed in the previous section. 
The ratios extracted from p+p data at the same energy are plotted for comparison (solid stars). 
The \pt coverage depends
on the pseudo-rapidity bins and extends up to \pteq{4} for $\eta = 2.6$ and $3.1$. 
At low \pt ($ < 1.5$ GeV/c) the \ptopip ratios exhibit
a rising trend with a weak dependence on centrality. The
dependency on centrality begins above $1.5$ GeV/c.
The ratios appear to reach a
maximum value at \pt around 2.5 GeV/c (whenever there is enough \pt coverage). 
The maxima of the ratios increase with the level
of centrality and at $\eta = 3.1$, are equal to about 2.5 and 1.5
for the $0-10\%$ and $40-80\%$ centrality bins, respectively. 
      The p+p ratios are consistent with Au+Au data at low \pt and begin to deviate 
significantly above \pteq{1}. At $\eta = 3.1$ a maximum value of the ratio
 of 0.55 is reached in p+p collisions which is
a factor of 4.5 smaller than that observed for central Au+Au reactions.

The values of the \aptopi ratios plotted in Fig. \ref{fig1_2} (right) are significantly
lower than the \ptopip ratios (note the difference in the vertical
scale), however, the centrality dependence shows the same features
as those observed in the \ptopip ratios, namely,
that the ratios for different centrality classes
are consistent with each other up to \ptapp{1.2} and
a strong dependence on centrality appears at larger transverse momenta  
reaching a maxmimum at similar \pt as the positive particles.
Looking at the p+p data alone, one can note the difference 
in shape between the  \ptopip and \aptopi ratios: a clear shift of the \aptopi
peaks towards lower \ptn, as well as a much broader \ptopip peaks. 
These large difference between the Au+Au and p+p
both in shape and overall 
magnitude may reflect significant medium effects in Au+Au at
\rootsnn{200} in the pseudo-rapidity intervals covered.

\begin{figure}[t]
\begin{minipage}[t]{63mm}
\centering
\includegraphics[width=1.03\textwidth,totalheight=0.3\textheight]
                {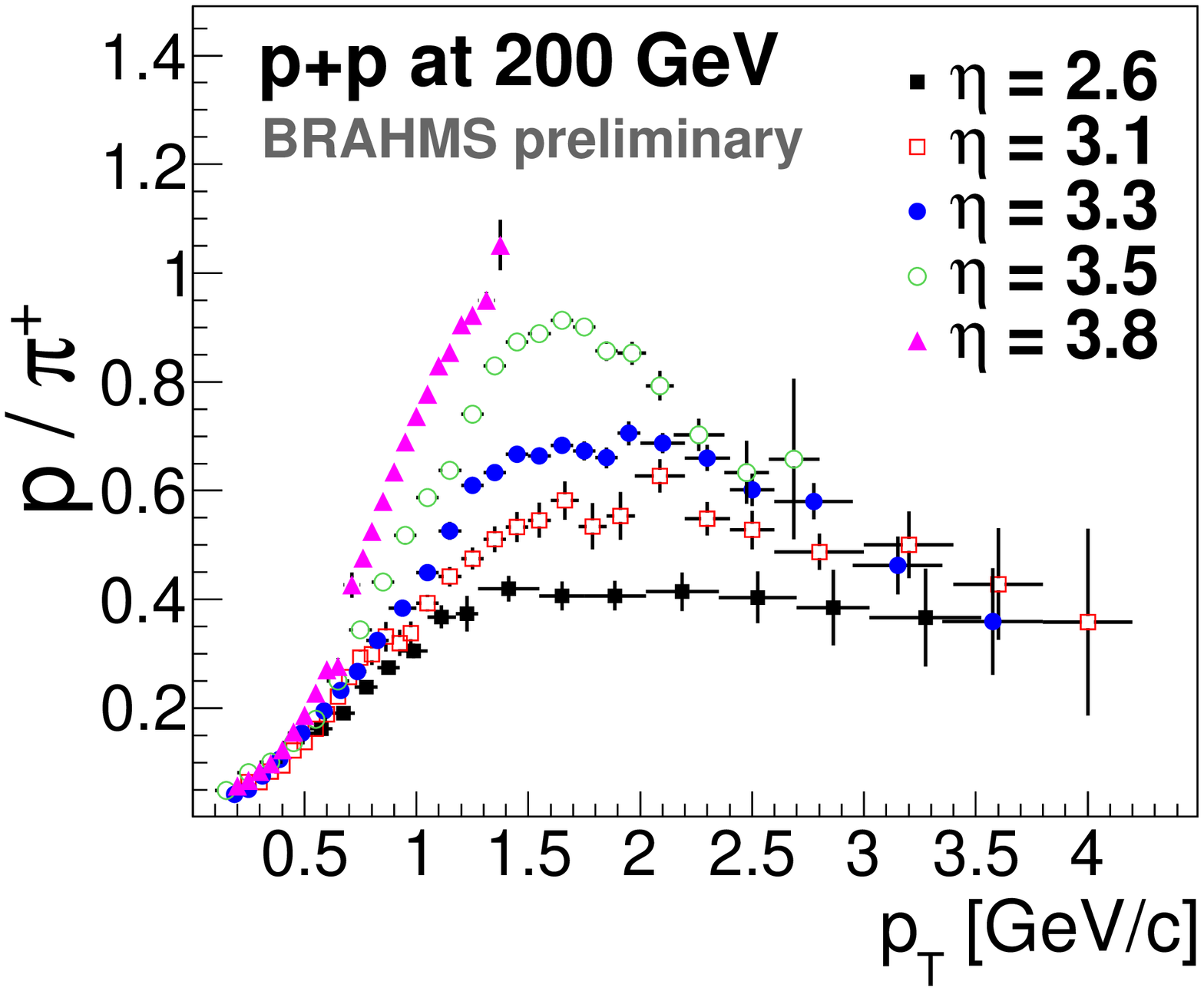}

\vspace{-0.4cm}

\caption{The \ptopi ratios at forward rapidities ($2.6 < \eta < 3.8$) 
in p+p at \roots{200}.}
\label{fig3}
\end{minipage}
\hspace{\fill}
\begin{minipage}[t]{63mm}
\centering
\includegraphics[width=1.03\textwidth,totalheight=0.3\textheight]
                {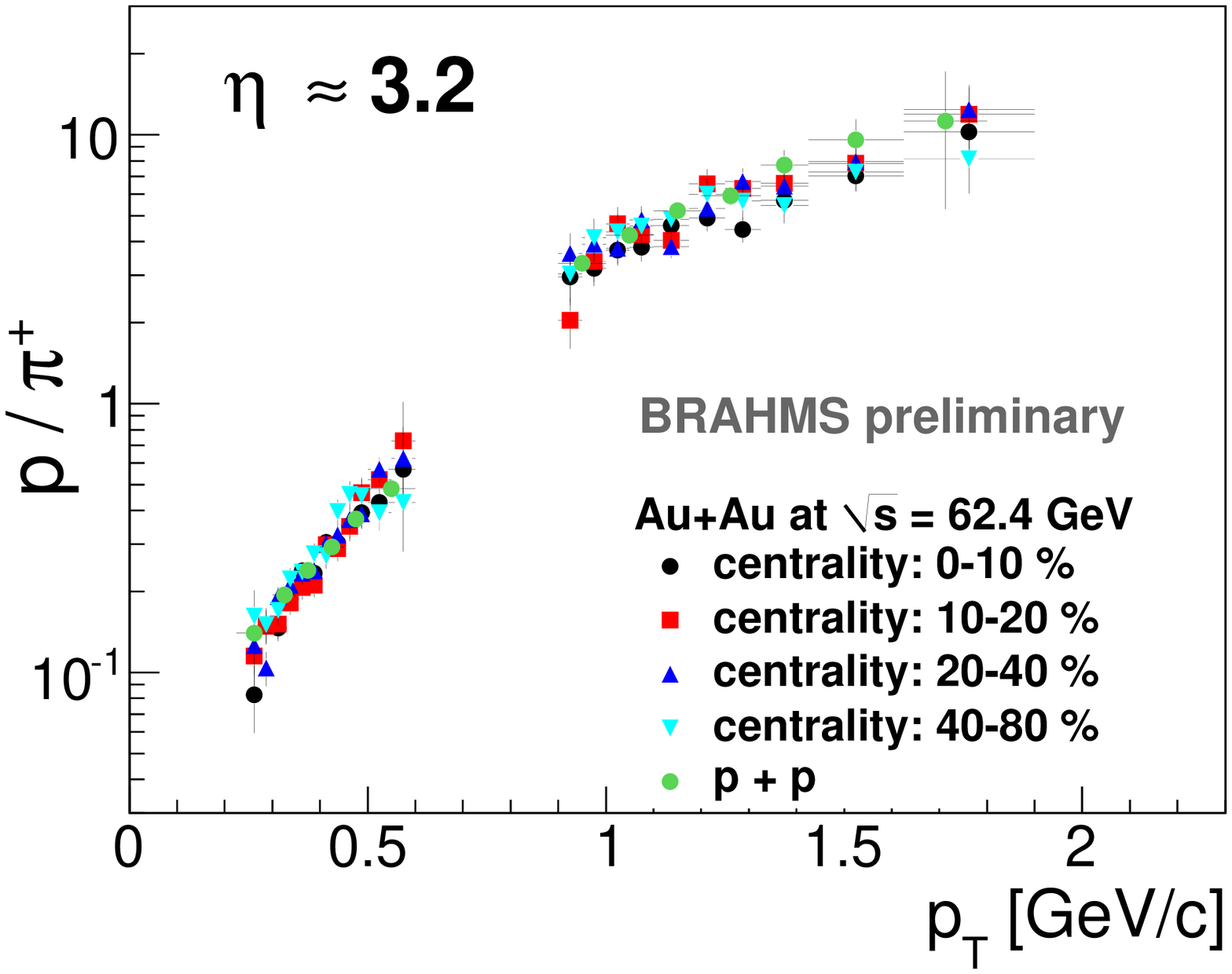}

\vspace{-0.4cm}

\caption{\ptopip ratio from p+p and Au+Au collisions
at $\sqrt{s_{NN}}$ = 62.4 GeV and $\eta \approx 3.2$.
}
\label{fig7}
\end{minipage}
\end{figure}

Fig. \ref{fig3} shows the \ptopip ratios as function of \pt in 
the pseudo-rapidity range
$2.6 < \eta < 3.8$ extracted from p+p reactions at \roots{200}. A very clear 
difference is found as the pseudo-rapidity changes from  
 $\eta=2.6$ to $\eta = 3.8$. But at high \pt all these ratios tend towards a common value 
 of about 0.4  consistent
with pQCD predictions \cite{hirano_PRC}.
Fig. \ref{fig7} compares the \ptopip ratio from p+p and Au+Au collisions
at $\sqrt{s_{NN}}$ = 62.4 GeV and $\eta \approx 3.2$ ($\mu_{B} \approx $ 250
MeV, \cite{ratio62}).
There is remarkable little difference in the \ptopip ratios over a
very wide range of the colliding system volume, namely, from p+p
reactions up to central Au+Au collisions.
It should be noted that (what is now shown) 
at $\eta$ = 2.67 the \ptopip ratio from central Au+Au reactions is
enhanced in respect to p+p collisions by a factor of 1.6, wheres
at $\eta$ = 3.5 the situation is reversed, namely, \ptopip ratio in
p+p exceeds that measured in central Au+Au by a factor of about 1.4. 
This indicates that the consistency observed at 
$\eta \approx 3.2$ is a results of simultaneous crossing of the ratios
for differnt systems (from p+p up to central Au+Au) at this
particular pseudorapidity bin.

  
We presented the $p_{T}$ dependence of the \ptopi ratios measured in Au+Au and p+p collisions
at energies 62.4 and 200 GeV as a function of pseudo-rapidity and collision
centrality (Au+Au). The data provide
the opportunity for studying baryon-to-meson production over a wide range
of the baryo-chemical potential, $\mu_{B}$. 
For Au+Au and p+p reactions at \rootsnn{200} the \ptopip and \aptopi
ratios show noticeable dependency on centrality at intermediate \pt
with a rising trend from p+p to central Au+Au collisions. We have
shown that \ptopip ratios are remarkably similar for central Au+Au at
\rootsnn{200}, $\eta \approx 2.2$ and central Au+Au at
\rootsnn{62.4},  $\eta \approx 0$, where the bulk medium is
characterized by the same value of $\bar{p}/p$ \cite{fv_qm09}. This observation,
together with the observed centrality dependence suggests, that at
these energies and pseudo-rapidity intervals,  particle production at
intermediate \pt is governed by the size and chemical properties of the
created medium. 
Finally,  the Au+Au and p+p measurements at
\rootsnn{62.4} shows that the \ptopip ratios for p+p and for all analysed Au+Au
centralities cross simultaneously at the same $\eta$
value ($\approx$ 3.2) and are consistent with each other in the covered \pt range e.g. from
0.3~GeV/c up to 1.8~GeV/c.

\end{document}